# A Fuzzy System based Approach to Extend Network Lifetime for En-Route Filtering Schemes in WSNs

M.K. Shahzad, *Member, IEEE*, L. Nkenyereye, *Member, IEEE*, S. M. Riazul Islam, *Member, IEEE*

[1]Authors' Affiliations

*Abstract*—Wireless sensor networks suffer from false report injection attacks. This results in energy drain over sensor nodes on the event traversal route. Novel en-route filtering schemes counter this problem by filtering these attacks on designated verification nodes. However, these filtering schemes among other limitations inherently are network lifetime inefficient. Generally, report traversal paths and verification nodes are also fixed. In this paper, we cater these limitations in our proposed scheme. Simulation experiments results show that proposed schemes outperforms existing en-route filtering schemes in networks lifetime. We employed a Fuzzy Logic System to select forwarding nodes from candidate nodes based on current network conditions. Proposed scheme gains in network lifetime, and energy-efficiency while having comparable false report filtering efficiency.

## 1. INTRODUCTION

Wireless sensor networks (WSNs) have witnessed mushroom growth due to recent advances in microelectromechanical systems (MEMS) [1]. En-route filtering schemes [2-7] have demonstrated several limitations; fix path routing, energy-hole problem, and often underlying routing [8] is not suitable for WSNs. Energy-hole problem in static sink based WSNs is caused by fasted rate of energy consumption around based station and critical paths. This results in network partitioning and deteriorated network lifetime.

In WSNs, false report injection attacks [9] are prevalent. An adversary with an intent to drain the scarce of energy of sensor nodes generate false reports from compromised node(s) and forward it to the base station. In order to counter these attacks, en-route filtering schemes use filtering nodes on the pre-determined path(s). Since, path and filtering nodes are fixed, making it vulnerable. In general, in these schemes, underlying routing is based on greedy or shorted path routing which does not perform well with regard to network lifetime.

Yet, another limitation of en-route filtering schemes is energy-hole problem [10]. This problem occurs since in static sink based WSNs, the rate of energy consumption around sink and on critical paths is faster as compare of other nodes and paths. This results in network partition(s) and severely effects network lifetime. In [11], a light-weight routing was presented to safe energy in sensor networks. In this work we use energy consumption model widely used [12]. Energy consumption values are taken from [13] assuming that underlying sensing platforms are Mica2 sensor nodes [14].

---

[1] MKS (mkshahzad@ieee.org) is with IEEE Seoul Section, LN (e-mail: nkenyele@sejong.ac.kr) and SMRI (e-mail: riaz@sejong.ac.kr) are with Sejong University, Seoul, South Korea.

In order to extend network lifetime without compromising security, we propose fuzzy based dynamic en-route filtering. In this approach, since the paths are not fixed communication loads are routed to several paths resulting in network lifetime extension. In the proposed scheme, dynamic path routing, there can exit several path since each time forwarding node is selected based on current energy level of candidate nodes, in addition to message authentication codes (MACs) and attacks frequency. In traditional schemes, attacks response is based on probabilistic method; independent of attacks intensity, security response is fix. However, in real scenarios, attacks frequency does vary with time. Therefore, a countermeasure which respond to current attack frequency is proposed in this paper. Main contribution of proposed scheme against existing schemes among other are:

- Dynamic path routing.
- Countermeasure against false report attacks based on current attacks frequency.
- Fuzzy Logic System based forwarding node selection.
- Avoiding energy-hole problem.

## 2. METHODOLOGY AND MODELS

In this section, experimental detail, assumptions, initialization, route set-up method, and fuzzy logic system to select forwarding nodes, and false report information method without causing extra energy or messages on sensor networks are presented.

### 2.1. Experimental Setup

Simulations experiments were conducted in a sensor field of area (500×500) m$^2$ that composed of three setups of randomly deployed 400, 700, and 1000 sensor nodes. The performance is evaluation in a custom built simulator using C/C++ in Microsoft Visual Studio 2010. All sensor nodes have initial 1 joules of energy and a circular range of 50m. The attack frequency or false traffic ratio (FTR) is set at 50%. In order to transmit one bit, 50nJ energy is required by electrical circuit while 100pJ power is needed by amplifier. The simulation experiment parameters are shown in Table 1.

### 2.2. Assumptions

The base station (*BS*) is aware of sensor nodes locations, keys, and node *IDs*. Energy consumption by radio components is calculated using first order radio model [12]. We assume Mica2 sensor mote as underlying sensing platform [13]. Energy consumption values for Mica2 sensing platforms are used from [14].

**Table 1:** Parameters

| Parameters | Values |
|---|---|
| Sensor | 1000 |
| Sensor field size | 500 x 500 m² |
| BS type | Static |
| BS location | (250, 0) m |
| $R_i$ | 50 m |
| Cluster h/w | 50 m |
| $E_{elec}$ | 50 nJ/bit |
| $E_{amp}$ | 100 pJ/bit/m² |
| Node energy | 1 Joules |
| Data packet | 32 bytes |
| Round | 128 bytes |
| Path loss constant ($\lambda$) | 2 |

## 2.3 Initialization

The *N* sensor nodes were randomly deployed within squared sensor field area $A^2$. The roughly same number of sensor nodes are positioned in each cluster. At initialization, each node is assigned a random location and plain text key. The *BS* initiate the query to find an event (i.e., fire/movement) in an area of interested.

## 2.4. Route Setup

In order to inquire an event in an area of interest, a request message is sent in a related cluster. In response, the *BS* creates a path from the *CH* in that cluster by using a fitness function using total energy(*TE*), node distance (*ND*) from BS and attack frequency (*AF*). The nodes among neighboring nodes with highest fitness value is selected a forwarding nodes. This process is continued till a path is created from *CH* to the *BS*. Later on, this function can be used to selected best paths when several paths are available. The fitness function is given in Equation (1).

$$F_v = MACs \times AF + (DN + TE)n; \text{ where } (1 < m, n < 0) \tag{1}$$

The Equation (1) parameters are used in fuzzy logic system as input fuzzy membership functions as explained in next section.

## 2.5. Fuzzy Logic System

In order to select next forwarding nodes to create a path, we employ fuzzy logic system shown in Figure 1. We define three fuzzy membership functions for three inputs DN, TE, and AF. Fourth membership function output fitness value (FV) which gives next or forward nodes with highest value, creating path by repetitions, based on fuzzy logic system. Fuzzy sets of each membership function have standard boundary value. Each fuzzy membership function have several members defining different values. For examples; FV membership function have low (L), and neutral (N), and high (H) labels. Therefore, in this Fuzzy system each input variable have three labels with total 24 rules (i.e., 2 × 3 × 4). Some selected rules are shown in Table 2. An evaluation function shown in equation (1) computes the fitness value ($F_v$) of each route.

The path with highest $F_v$ is selected as data forwarding path. As underlying network conditions such as energy of individual nodes on a path changes, the preferred at a given time changes as well. Making it dynamic path routing which consider current network conditions as a result balancing communication loads over larger group of nodes or paths. This helps in avoiding energy-hole and increasing network lifetime.

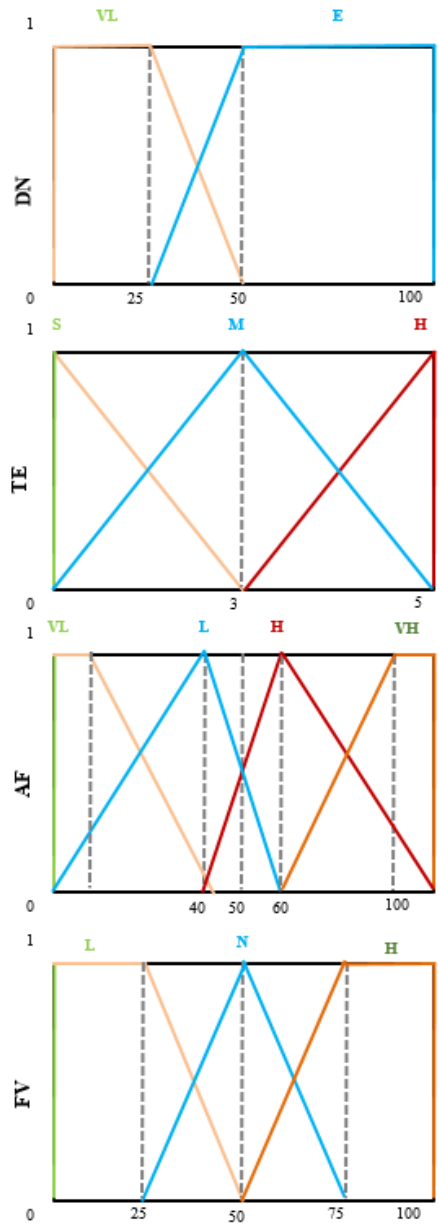

**Figure 1.** Fuzzy membership functions for fitness value evaluation to select forwarding nodes

**Table 2.** Fuzzy if-then rules

| Rule No. | IF | | | THEN |
|---|---|---|---|---|
| | TE | ND | AF | FV |
| 1 | VL | S | VL | L |
| 3 | VL | S | H | N |
| 12 | VL | H | VH | N |
| 15 | E | S | H | N |
| 17 | E | M | VL | L |
| 19 | E | M | H | N |
| 23 | E | H | H | H |

### 2.6. False reports Attacks Models

Our communication model is based on query-repose model. For one session, expected number of events are known. After a legitimate report is received at the BSthe corresponding counter is incremented by one. For this case no messages or energy is required to count total number of reports. Fabricated reports can be dropped intermediate nodes or on BS. In case dropping on an intermediate node the BS after a time window will know a report is dropped and when it is dropped on the BS, it also knows it. Therefore, by using following function, we can determine current attacks frequency or false traffic ratio without causing extra messages or energy consumption as illustrated in Equation (2).

$$AF = \sum_{e=1}^{n} \frac{F_R}{F_R + L_R} \qquad (2)$$

## 3. EVALUATION

In this section, performance analysis and results are presented.

### 3.1 Performance Analysis

In order to evaluate the performance analyses, we use first node depleted (FND) and last node depleted (LND) performance metrics to measure the network lifetime of proposed and existing schemes. For security, we measure filtering efficiency with filtering ratios of these schemes. Fuzzy based dynamic path selection help spread communication loads over multiple paths and larger group of sensor nodes. Essentially, this helps balancing energy consumption and thus avoiding energy-hole problem. This effectively can increase network lifetime as a result.

Security response in proposed scheme is based on underlying network conditions, in particular currently attack frequency. In case of high attack frequency, energy is saved as proposed method select more secure path resulting in dropping more number of false report en-route. In case of low frequency of false report attacks, a path with less number of verification noes is selected thus legitimate reports have to go through less number of verification consequently saving energy.

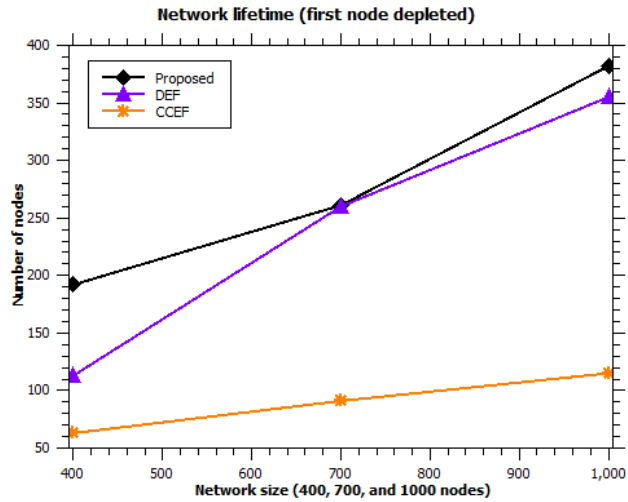

**Figure 2.** Network lifetime using first node depleted metric.

Thus our proposed method not only saves energy but extends network lifetime while maintaining similar filtering efficiency as compare to the existing schemes.

### 3.2 Performance Results

In this section, performance results for network lifetime and filtering efficiency are presented.

The network lifetime analysis is shown using FND and LND in Figures 2 and 3 respectively. In Figure. 2, our proposed scheme shows a significant improvement of 3.104 times of CCEF and 1.147 folds over DEF schemes. For LND performance metric, proposed schemes gains 2.545 times over CCEF and 1.055 folds over DEF scheme respectively as illustrated in Figure 3.

In order to measure filtering efficiency, the false report injection attacks detected are divided by total number of false and legitimate reports. The filtering efficiency for 50% false report injection attacks is shown in Figure 4. The filtering capacity of CCEF is 98.6%, DEF is 99.3%, and proposed scheme is 99.4%. It can be observed that our proposed scheme gives similar filtering efficiency while having significant gains in network lifetime.

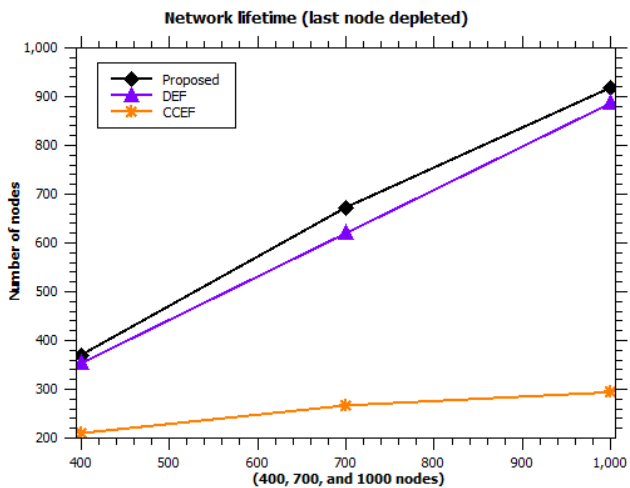

**Figure 3.** Network lifetime using last node depleted metric.

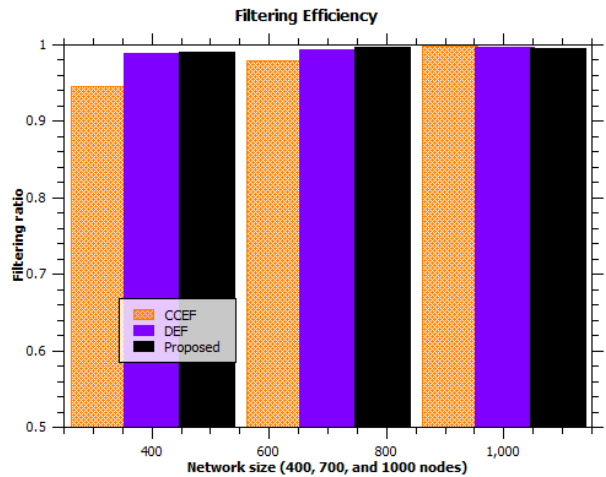

**Figure 4.** Filtering efficiency.

## 4. CONCLUSIONS AND FUTURE WORKS

In this paper we proposed fuzzy logic system based scheme which extended network lifetime in en-route filtering schemes without compromising security. In future, we will perform more experiments and include energy-efficiency results. Moreover, we may include re-clustering to get further extension in network lifetime.

### Acknowledgement
This research was supported by the Sejong University New Faculty Program through the National Research Foundation of Korea.